\documentclass[twocolumn,showpacs,preprintnumbers,amsmath,amssymb]{revtex4}

\usepackage{graphicx}
\usepackage{dcolumn}
\usepackage{bm}

\begin{document}

\preprint{arxiv}

\title{Quantitative model for anisotropy and reorientation thickness of the magnetic moment
in thin epitaxial strained metal films}

\author{Artur Braun}
\altaffiliation[present address:]
{Empa\\Swiss Federal Laboratories for Materials Science and Technology\\Laboratory for High Performance Ceramics\\Ueberlandstrasse 129\\CH-8600 Duebendorf\\Switzerland}
\email{artur.braun@alumni.ethz.ch}
\affiliation{Environmental Energy Technologies\\
Ernest Orlando Lawrence Berkeley National Laboratory\\One Cyclotron Road, MS 70-108 B\\
Berkeley, CA 94720, USA
}

\homepage{http:// http://www.empa.ch/plugin/template/empa/613/*/---/l=1}

\date{\today}

\begin{abstract}
A quantitative mathematical model for the critical thickness of strained epitaxial metal films is presented,
at which the magnetic moment experiences a reorientation from in-plane to perpendicular magnetic anisotropy.
The model is based on the minimum of the magnetic anisotropy energy with respect to the orientation
of the magnetic moment of the film.
Magnetic anisotropy energies are taken as the sum of shape anisotropy, magnetocrystalline anisotropy and
magnetoelastic anisotropy, the two latter ones being present as constant surface and variable volume
contributions. Apart from anisotropy materials constants, readily available from literature,
only information about the strain in the films for the determination of the magnetoelastic anisotropy
energy is required.
Application of continuum elasticity theory allows to express the strain in the film in terms of substrate
lattice constant and film lattice parameter, and thus to obtain an approximative closed expression
for the reorientation thickness in terms of lattice mismatch.
The model is successfully applied to predict the critical thickness with surprising accuracy. 
\end{abstract}

\maketitle

\section{\label{sec:level1}Introduction}

The efficiency of magnetic data storage media depends critically on the presence of a perpendicular
magnetization, this is, the easy magnetic axis is perpendicular to the surface of the magnet in the absence
of an external magnetic field.
Generally, the magnetization lies in some preferred direction with respect to the crystalline axes or
to the external shape of the magnet - this is known as magnetic anisotropy.
The energy involved in rotating the magnetization from a direction of low energy (easy axis) towards one of
high energy (hard axis) is typically of the order 10$^{-6}$ to 10$^{-3}$ eV/atom and thus a very small
correction to the total magnetic energy \cite{Bruno}.   
The physical origin for the anisotropy is the symmetry breaking of the rotational invariance of the dipole-dipole
interaction and the spin-orbit coupling with respect to the spin quantization axis.         
Atoms at interfaces in ultrathin films or multilayers show a much stronger anisotropy than atoms in the
bulk. Also, atoms in strained lattices show a stronger anisotropy.

While it has been an established fact for many years that epitaxial strain in ultrathin metal films can have
significant impact on the stabilization of perpendicular magnetization \cite{Braun,4,48,1}, many relevant
publications still ignore magnetoelasticity for the interpretation of magnetic anisotropy data.                    

The relationship between the magnetic anisotropy and elastic strain is also important in structural geology.
Elastic strain is a marker that permits to draw conclusions about deformation history of rocks.
Since such strain data are not always directly available, quantitative modeling helps
to retrieve strain data from magnetic anisotropy data \cite{Jezek}.
Reviews about the interplay between mechanical stress and magnetic anisotropy in thin films and in tectonics are given by
Sander \cite{Sander}, and by Borradaile and Henry \cite{Borradaile}, respectively.

In the next section, the specific contributions to magnetic
anisotropy energy will be given and summarized.  The magnetoelastic
energy will be expressed in terms of lattice spacings in film and
substrate in order to account for the strain in films.
The derivative of the total magnetic anisotropy with
respect to the film crystal axes, i.e. the minimum of this energy will
yield a thickness at which the magnetization switches from in-plane to
out-of-plane.
The terms {\em out-of-plane} and {\em perpendicular} are used equivocally in this manuscript.
The Bain path is a functional relationship between in-plane and out-of-plane
lattice distances of a bulk metal phase and its corresponding tetragonally distorted
equilibrium phases under epitaxial stress.
It can be regarded as a sequence of tetragonal states produced by epitaxial
stress on an equilibrium tetragonal phase \cite{6,Alippi,Marcus2} and
serves as a criterion for the identification of equilibria phases of
strained structures and for the interpretation of LEED data \cite{Feldmann}.
The concept is being improved still to date and allows, for instance,
deeper insight in phase transitions during epitaxial growth \cite{Marcus2}, or just the conversion
of film thickness data from Monolayers to Angstrom \cite{BraunBain}.
For the present model, the Bain path plays a key role.
To the best of the authors knowledge, the present model is the first of its kind which gives a quantitative
prediction of the critical reorientation thickness.

\section{Formulation of the model}
\subsection{\label{sec:level2}Magnetocrystalline anisotropy energy}

The magnetocrystalline anisotropy energy G$_{cryst}^V$($\Omega_M$) is written as expansion of the
cosine of directions $\alpha_1$, $\alpha_2$, and $\alpha_3$ of $\Omega_M$:
\begin{eqnarray}
G_{cryst}^V(\Omega_M)=b_0(H_M) + \sum_{i,j} b_{i,j}(H_M)\alpha_i \alpha_j \nonumber\\
 + 
\sum_{i,j,k.l} b_{i,j,k,l}(H_M)\alpha_i \alpha_j \alpha_k \alpha_l + \ldots .
\end{eqnarray}
$\Omega_M$ is the unit vector of magnetization direction with components $\alpha_1$, $\alpha_2$, $\alpha_3$
or the polar angles $\theta$ and $\phi$.
For crystals with cubic symmetry, the expression in Eq. (1) simplifies to
\begin{eqnarray}
G_{cryst}^V(\Omega_M) = K_0 + K_1(\alpha_1^2 \alpha_2^2+\alpha_2^2 \alpha_3^2+\alpha_3^2 \alpha_1^2)  \nonumber\\
+ K_2 \alpha_1^2 \alpha_2^2 \alpha_3^2 
+ K_3(\alpha_1^2 \alpha_2^2+\alpha_2^2 \alpha_3^2+\alpha_3^2 \alpha_1^2) + \ldots .
\end{eqnarray}
with the coordinate system being in line with the crystal axes, and:
$\alpha_1$=$\sin\theta\cos\phi$, $\alpha_2$=$\sin\theta\sin\phi$, and $\alpha_3$=$\cos\theta$.
The vector of magnetization, {\bf M}, includes with the surface normal of the (001) plane the angle $\theta$
and with the [010]-direction the angle $\phi$. The expression in Eq. (2) converges well and thus permits to
interrupt the expansion after the third order. K$_0$, K$_1$ etc. are the magnetocrystalline
anisotropy constants. Note, that these {\em constants} are temperature dependent materials parameters.
But here we will treat the case for ambient temperature only.

Eq. (2) describes the magnetocrystalline
anisotropy energy or a single atom of one monolayer in the volume of the film. This contribution grows 
proportional with the film thickness.
Atoms at surfaces and interfaces are not described properly by Eq. (2) anymore.
Because of the reduction of symmetry at surfaces and interfaces, second order terms come into play.
For a surface with tetragonal symmetry, such as the cubic (001) plane, the surface magnetocrystalline
of a single atom is
\begin{eqnarray}
G_{cryst}^S (\Omega_M)~=~K^S_1 \sin^2 \theta  \nonumber\\
~+~(K^S_2 + K^{S'}_2 \cos(4\phi))\sin^4 \theta + \ldots .
\end{eqnarray}

\subsection{Shape anisotropy energy}
The shape anisotropy energy
\begin{equation}
G_{shape}(\Omega_M) = -\frac{1}{2}\int_V dV {\bf M}({\bf r}) \cdot {\bf H}_d({\bf r}) \; ,
\end{equation}
with demagnetization field {\bf H}$_d$(r) and magnetization {\bf M}({\bf r}), depends on the external
shape and geometry of the magnet.
Ultrathin films can be well approximated by a plate with infinite lateral extension, such as:
\begin{equation}
G_{shape}^V(\theta) = K_{shape}^V \sin^2 \theta \; .
\end{equation}
Eq. (5) describes the shape anisotropy energy of a single atom in a monolayer of such a film.
Considering infinitesimally thin slices yield generally the same angular dependence:
\begin{equation}
G_{shape}^S(\theta) = K_{shape}^S \sin^2 \theta \; .
\end{equation}
K$_{shape}^V$ and K$_{shape}^S$ are the corresponding anisotropy energy constants.

\subsection{Magnetoelastic anisotropy energy}
For the strained lattice of a magnetized body, the energy terms may depend on the strain tensor
{\bf $\epsilon$} and on $\Omega_M$; this is the magneto-elastic energy.
For small lattice mismatches, the energy terms can be expanded in spherical harmonics and in
powers of {\bf $\epsilon$}:
\begin{equation}
G_{magn.el}(\Omega_M, \epsilon) = \sum_{i,j,k,l} B_{i,j,k,l} \epsilon_{i,j}\alpha_i \alpha_j + \ldots .
\end{equation}
The crystalline symmetry manifests itself in a coupling of the expansion coefficients.
For a cubic system, the expression reads
\begin{eqnarray} 
G_{magn.el}(\Omega_M, \epsilon) = \nonumber\\
B_1(\epsilon_{11}\alpha_1^2+\epsilon_{22}\alpha_2^2 +\epsilon_{33}\alpha_3^2) \nonumber\\
+~2 B_2(\epsilon_{12}\alpha_1\alpha_2 
+\epsilon_{23}\alpha_2\alpha_3 +\epsilon_{31}\alpha_3\alpha_1) + \ldots .
\end{eqnarray}
The $\epsilon_{ii}$ describe the strain and the $\epsilon_{ij}$ describe the shear deformation.

We now want to derive an expression for the magneto-elastic anisotropy energy in terms of lattice deformation.
Let us consider here a face centered cubic lattice.
The nearest neighbor (NN) distance of the atoms in a cubic lattice with no strain be $a_0$.
For the lattice under strain, the NN distance in the tetragonal plane be $a$.
For pseudomorphic growth, the latter value is the NN distance of the atoms in the substrate.
The expansion is equal to the lattice mismatch:
\begin{equation}
\epsilon_{11} = \epsilon_{22} = f =\frac{a-a_0}{a_0} \; .
\end{equation}

For the contraction, we find

\begin{equation}
\epsilon_{33} = \frac{c-a_0\sqrt{2}}{a_0\sqrt{2}} = \frac{c}{a_0\sqrt{2}}-1 \; .
\end{equation}

The variation of $c$ with $a$ is given by the Bain path, this is \cite{6}

\begin{equation}
\frac{c}{a_0} = \left(\frac{a_0}{a}\right)^{\gamma} \; , c_0 = a_0\sqrt{2}
\end{equation}
so that we may write
\begin{equation}
\epsilon_{33} = \left(\frac{a_0}{a}\right)^{\gamma}-1 \; , \gamma=2\frac{c_{12}}{c_{11}} \; .
\end{equation}
Here, $c_{11}$ and $c_{12}$ are the elastic constants for the corresponding film material.
Some elastic constants and lattice parameters for Fe, Co, and Ni are listed in Table I.

\begin{center}
\begin{tabular}{l l  l l l}
\hline
\hline
Element   & T(K) & c$_{11}$ [GPa] & c$_{12}$ [GPa]  & a [\AA] \\
\hline 
bcc Fe  & 520 & 230 & 135  & 2.87 \\
fcc Fe  & 1428 & 154 & 122  & 3.65 \\
\hline
fcc Co  & 300 & 242 & 160  & 3.54 \\
hcp Co  & 300 & 307 & 165 & 2.82\\
\hline
fcc Ni  & 300 & 247 & 153  & 3.524 \\
bcc Ni  & 300 & - & -  & 2.78 \\
\hline
\hline
\end{tabular}
\end{center}
Table I: Elastic constants and lattice parameters for some Fe, Co, and Ni phases \cite{72,73},
hcp Co after \cite{Gump}.

We neglect shear effects in the distorted films and insert the expressions from Eqs. (9)-(11) in Eq. (8)
and obtain in first order approximation
\begin{eqnarray}
G_{magn.el}(a,a_0,\theta) = \nonumber\\
B_1 \left \lbrace \left(\frac{a}{a_0}-1\right)\sin^2\theta+
\left(\left(\frac{a_0}{a}\right)^{\gamma}-1 \right) \cos^2\theta \right \rbrace \; .
\end{eqnarray}

\subsection{Summary of contributions}
All contributions considered so far contain bulk- and surface/interface contributions, except for
the magnetoelastic anisotropy energy, for which no surface- or interface contributions were
available from literature.
For a film with $d$ monolayer thickness, the anisotropy energy per unit surface is thus

\begin{equation}
G = d \cdot \left( G_{cryst}^V + G_{shape}^V+ G_{magn.el}^V \right) + G_{cryst}^S + G_{shape}^S
\end{equation}
With the exception of $G_{cryst}^V$, we insert the expressions for the particular contributions
in Eq. (14) and obtain

\begin{widetext}
\begin{eqnarray}
G(a,a_0,\theta) = \left( K_1^S+K_{shape}^S\right)\sin^2\theta 
+ d \cdot \left(K_{shape}^V \sin^2 \theta + B_1 \left
\lbrace \left(\frac{a}{a_0}-1\right) \sin^2 \theta + 
\left(\left(\frac{a_0}{a}\right)^{\gamma}-1\right) \cos^2 \theta \right \rbrace \right)
\end{eqnarray}
\end{widetext}
Omission of $G_{cryst}^V$ is justified for ultrathin films, because the error inferred by this contribution
is negligible. With increasing thickness, however, this contribution will dominate together with the shape anisotropy, also,
because magnetoelastic contributions will become smaller due to relief of strain in thikcer films.

We can replace $\cos^2\theta$ by $(1-\sin^2\theta)$ in Eq. (15), so that $G(a,a_0,\theta)$ becomes a function
linear in $\sin^2\theta$.
It is of interest now to know the conditions for which the angle $\theta$ becomes $0$.
Then,
the magnetic anisotropy energy has a minimum, and this state corresponds to perpendicular magnetization.
To determine, for which particular angles $\theta$ the magnetic anisotropy energy becomes a minimum,
we calculate the first derivative of $G(a,a_0,\theta)$ in Eq. (15) with respect to $\theta$.
The derivative vanishes for $\theta$=0$^{\circ}$ and 90$^{\circ}$.
For these cases we obtain an explicit expression for the critical thickness, at which we have a
reorientation of the magnetization:

\begin{equation}
d_{crit} = \frac{-\left( K_{shape}^S + K_{cryst}^S \right)}{K_{shape}^V+B_1\left(\frac{a}{a_0}-\frac{a}{a_0}^{-2\frac{c_{12}}{c_{11}}} \right)}
\end{equation}
The expression in Eq. (16) thus allows the prediction of the critical magnetic reorientation thickness,
within the limits and approximations made in the model.
More generally, Eq. (15) allows to determine the direction of magnetization in a thin film for any given set of parameters.

\section{Results and Discussion}
The model is applied to thin films of Fe, Co, and Ni, and compared with experimental data taken from
literature.
Structure and magnetism of these three metals have been subject of intense research for decades, c.f.
\cite{Sander,Heinz} and a whole body of experimental data is readily available in literature.

We begin with Figure 1, which shows a schematic diagram of the critical reorientation thickness $d_{crit}$ (Eq. 16)
as a function of $a/a_o$ for four different sets and signs of $K_S$ and $B$.
Phase ranges for perpendicular magnetization are marked with $\perp$; ranges of in-plane magnetization
are marked with $\parallel$.
The modulus for the anisotropy energies was chosen arbitrarily to $K^S_{cryst}$~=~4.0 x 10$^{-4}$,
$B$~=~0.20 x 10$^{-4}$, and $K^S_{shape}$~=~8.0 x 10$^{-4}$ eV/atom.
The ratio $a/a_o$ is a measure for the strain in the film.
For $a/a_o$ $<$ 1, the film in-plane lattice parameter is compressed with respect to the substrate.
For $a/a_o$ $>$ 1, it is expanded.
The regions of in-plane and out-of-plane magnetization are displayed depending on the lattice mismatch and
the sign of the anisotropy constants. $G^V_{cryst}$ was not taken into account here.

\begin{figure*}
\includegraphics{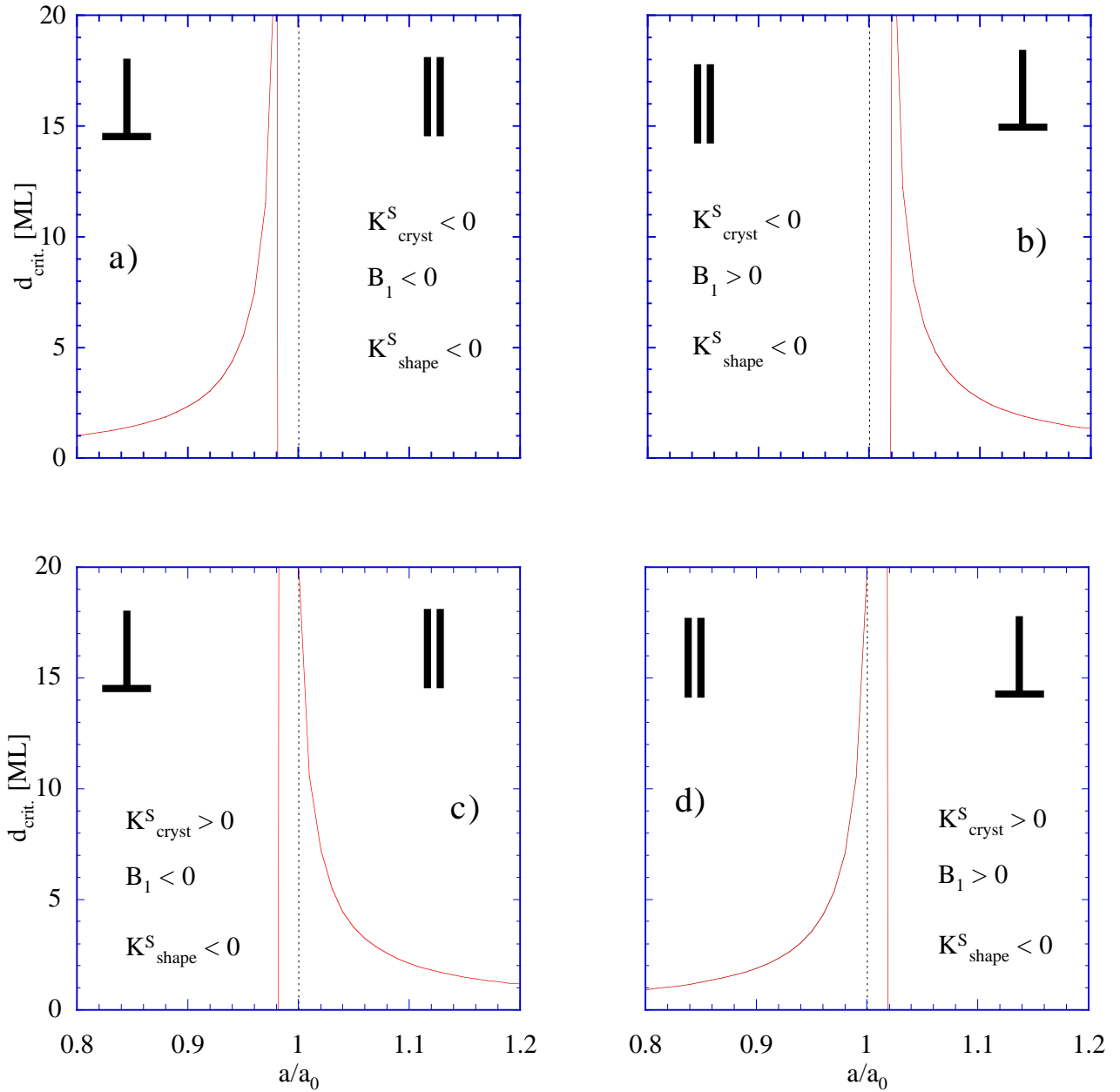}
\caption{\label{fig:wide}Schematic diagram of the critical reorientation thickness. Each sub-figure represents a specific case with
combinations of sign of magnetoelastic anisotropy constant $B_1$ and surface/interface anisotropy constant
$K_{cryst}^S$. Shape anisotropy is taken negative for all cases.}
\end{figure*}

For these systems, the magnetocrystalline anisotropy energy is dominant in terms of modulus.
The position a/a$_o$=1 denotes the relaxed state of the film and is marked by a dotted vertical line.
In addition, a solid vertical line marks the specific ratio a/a$_o$, beyond which a polar magnetization is not possible
anymore, regardless the film thickness.
This is because in Eq. (16), the shape anisotropy and magnetoelastic anisotropy energy cancel out
in the denominator.

P. Bruno \cite{Bruno} has compiled a number of anisotropy constants, displayed in Table II and III,
which are used in the present work.

\begin{center}
\begin{tabular}{l c c c }
\hline
\hline
      & bcc Fe   & hcp Co  & fcc Ni   \\
\hline
 
K$_1$  & 4.02 10$^{-6}$ (a) &5.33 10$^{-5}$ (b) & -8.63 10$^{-6}$ (a)\\
K$_2$  & 1.44 10$^{-8}$ (a) &7.31 10$^{-6}$ (b)& 3.95 10$^{-6}$ (a)\\
K$_3$  & 6.60 10$^{-9}$ (a)&    -    & 2.38 10$^{-7}$ (a)\\
K$_3$'  &      -       &  8.40 10$^{-7}$    (c)  & 6.90 10$^{-7}$ (c)\\
   &          &    &  \\
K$_{shape}^V$  & -8.86 10$^{-4}$ (c) &  -5.85 10$^{-4}$ (c)&  -0.74 10$^{-4}$ (c)\\
   &          &    &  \\
B$_1$  & -2.53 10$^{-4}$&  -5.63 10$^{-4}$ &  6.05 10$^{-4}$  \\
B$_2$  & 5.56 10$^{-4}$ &  -2.02 10$^{-3}$ &  6.97 10$^{-4}$  \\
B$_3$  &  - &  -1.96 10$^{-3}$ &  -   \\
B$_4$  &  - &  -2.05 10$^{-3}$ &   -  \\
\hline
\hline
\end{tabular}
\end{center}
Table II: Volume contributions of anisotropy energy constants in [eV/atom].  Data
for magnetocrystalline anisotropy energy are valid for T=4.2~K.
(a) \cite{Escudier}, (b) \cite{Rebouillat}, (c) \cite{Paige},
(e) \cite{Bruno}.
Magnetoelastic constants were obtained from \cite{Bruno} and \cite{Gersdorf,Hubert,Bower,Kimura}
and are valid for 297 K.

The magnetocrystalline anisotropy energy is of the order 10$^{-6}$ eV/atom and
is always dominated by the shape anisotropy energy by one (Ni) or two (Fe) order of magnitudes.
They have always a negative sign and thus favor magnetization in the film plane.
If we consider a Ni film with a lattice mismatch of $f$=1\%, the negative magnetoelastic anisotropy
energy of Ni will overcompensate the shape anisotropy energy by a factor of ten.
The surface and interface contributions to the anisotropy energy are constants that add up in the overall
anisotropy energy of the thin film.
Table III gives an overview of the surface/interface anisotropy constants of a number of systems studied in
the past, mostly one single crystalline substrates.
\begin{center}
\begin{tabular}{l c c  }
\hline
\hline
System & K$_S$ [mJ/m$^2$] & Reference \\
\hline 
Fe(001)/UHV & 0.96 & \cite{51}  \\
\hline
Fe(001)/Au & 0.47, 0.40, 0.54 & \cite{51} \\
\hline
Fe(001)/Cu & 0.62 & \cite{51} \\
\hline
Co/Au(111) & 0.42 & \cite{52} \\
\hline
Co/Cu(111) & 0.53 & \cite{53,48} \\
\hline
Co/Ni(111) & 0.31, 0.20, 0.22 & \cite{54,55,56} \\
\hline
Ni(111)/UHV & -0.48 & \cite{57} \\
\hline
Ni/Au(111) & -0.15 & \cite{58} \\
\hline
Ni(111)/Cu & -0.22, -0.3, -0.12 & \cite{58,59,60} \\
\hline
Ni/Cu(001) & -0.23 & \cite{60} \\
\hline
\hline
System & K$_{Shape}^S$ [erg/cm$^2$] & Reference \\
\hline
Fe(001)/UHV & -0.27 & \cite{61} \\
\hline
Fe/Ag(001) & -0.12 & \cite{61} \\
\hline
Ni(001)/UHV & -0.017 & \cite{62} \\
\hline
Ni/Cu(001) & -0.025 & \cite{62} \\
\hline
\end{tabular}
\end{center}
Table III: Surface contributions of anisotropy energy constants \cite{Bruno}.
Proper conversion yields the order of approximately 10$^{-4}$~eV/atom.

Ultrathin films often exhibit perpendicular magnetization up to several monolayers, which is stabilized
by the positive surface and interface anisotropy energy.
For Fe and Co, elastic strain works against this effect, and therefore perpendicular magnetization is not favored
anymore.
This is not the case for Ni. Here, we always register a negative magnetocrystalline surface/interface
anisotropy, and a perpendicular magnetization is favored.
Elastic strain, such as in the case of epitaxial growth, causes a positive contribution of the magnetoelastic
anisotropy energy and thus enhances this effect.
Since the magnetoelastic anisotropy energy is a volume contribution, this is, it increases to the same extent
as the volume or the thickness of the film, thick films of arbitrary thickness with perpendicular magnetization can be 
grown - at least theoretically.
The practical limitation is that thick films will not maintain the necessary lattice mismatch, i.e.
the elastic strain, and relief most of the elastic energy by dislocations \cite{Matthews2,Lambert}.

In Fig. 2, we show the variation of the critical thickness (curved lines) for Fe, Co, and Ni as a function of a/a$_o$.
Solid lines indicate $d_{crit}$ after Eq. (16), while dashed lines take also the volume contribution
of the magnetocrystalline anisotropy into account.
The vertical solid and dashed lines indicate the ranges where the denominator in Eq. (16) vanishes
and beyond which, according to the model, no perpendicular magnetization is possible.

\begin{figure*}
\includegraphics{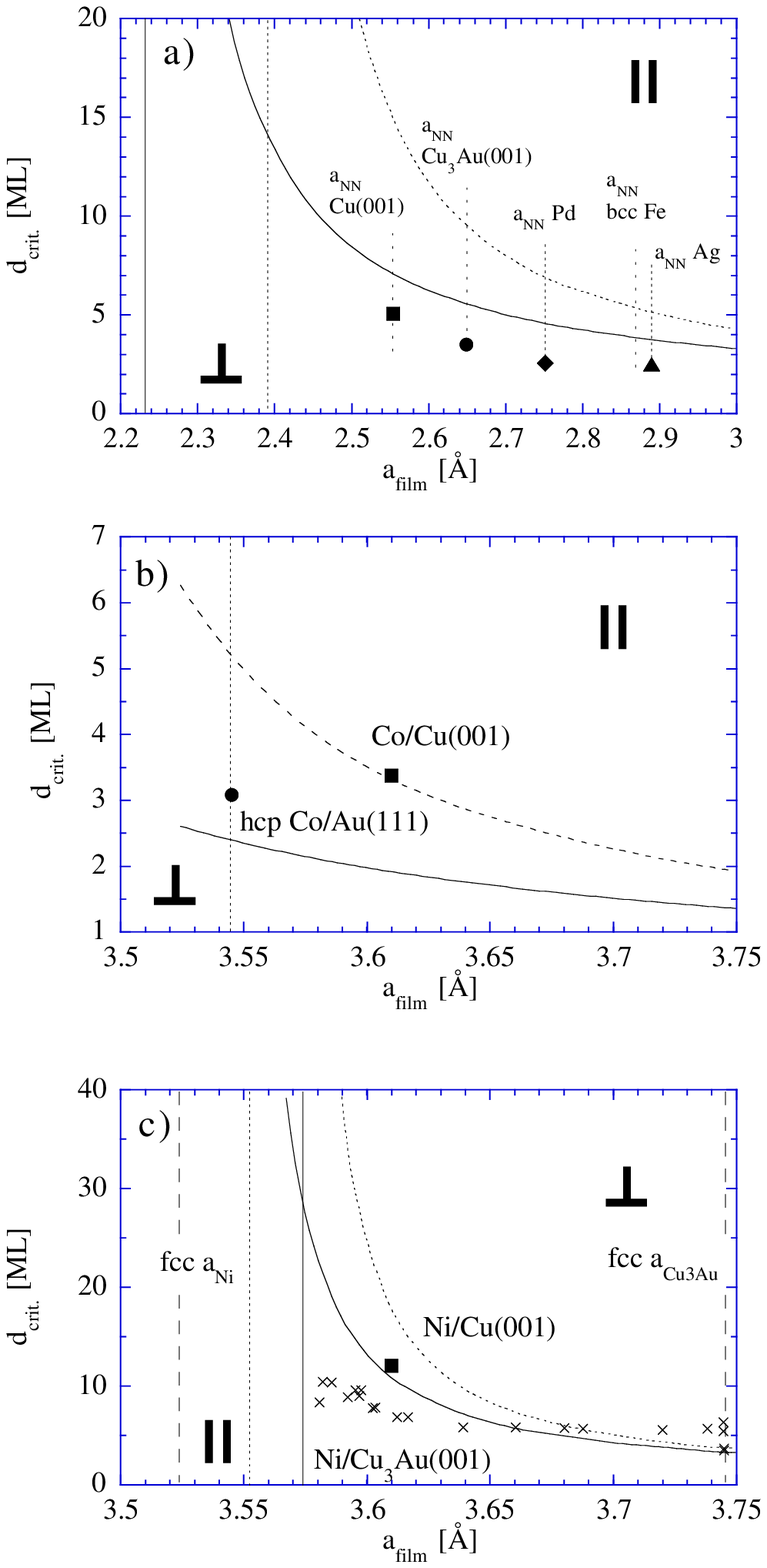}
\caption{\label{fig:wide}Critical thickness for Fe (plot a), Co (plot b), and Ni (plot c).}
\end{figure*}

First, we note that thin Fe films have a perpendicular magnetization ($\perp$).
Thick Fe films have the magnetization vector in the film plane ($\parallel$).
This is the situation as represented by Figure 1, plot c).
The critical thickness for the reversal of the magnetization vector is increasing with
decreasing strain in the film.
According to our model, particularly an ultrathin Fe film has a perpendicular magnetization.
The dotted line was obtained after Eq. (16).
The volume contribution of the magnetocrystalline anisotropy energy was not taken into account.
But we can estimate the influence of this contribution on the critical film thickness.
For a 10 ML thick film, the contribution does not exceed 4.02 10$^{-5}$ eV/atom.
The solid line in Fig. 2 takes this contribution into account and is thus an upper approximation for
the critical film thickness.

Figure2, plot a), contains four experimentally obtained values for the critical
reorientation thickness of bcc-Fe films,
i.e. Fe/Cu(001) \cite{Mueller}, Fe/Cu$_3$Au(001) \cite{Feldmann3}, Fe/Ag(001) \cite{Qiu},
and Fe/Pd(001) \cite{Feldmann}.
All of them are located below the curves for the calculated critical thickness.
For the system Fe/Cu(001) and Fe/Cu$_3$Au(001) it is believed that the magnetic properties depend on
the preparation conditions of the films, which could not be taken into account in the model.
In addition, for the latter case, interdiffusion of Au atoms in the Ni film
is believed to have changed the magnetic properties of the Ni film \cite{Feldmann}.
It also should be mentioned that the temperature dependence of the anisotropy constants was not taken
into account in the present model.
But it is striking that the experimentally obtained critical thickness increases
with decreasing film strain, as predicted by our model, i.e. experimental data and calculations show
the same trend.
One thorough confirmation of the clear separation between surface and volume contributions is given by Thomas et al.
\cite{Thomas}.
They found that the uniaxial magnetic anisotropy of {\em very thin} Fe/GaAs(001) is not a result of
magnetoelastic coupling nor a result of shape anisotropy, but a result of interface anisotropy.
On the other hand, they find that the further evolution of magnetic anisotropy with increasing Fe film thickness
is a result of the competition between magnetoelastic coupling and interface anisotropy.
It is worthwhile to mention that their work was based on a thorough X-ray structure analysis of the films.

In the middle plot in Fig. 2, plot b), the critical thicknesses for Co films with and without correction
for the volume contribution of the magnetocrystalline anisotropy are given.
Ultrathin Co films have an easy magnetic axis perpendicular to the film plane, and thick planes are
magnetized in the film plane.
This is a situation similar as with Fe films, as represented by Figure 1, plot c).
The experimental data are taken from Co/Au(111) \cite{48} and Co/Cu(001) \cite{Matthews}.
In the former case, elastic strain in the film is anticipated, but not quantified. The lattice mismatch for
this system would be theoretically about 14\%. But it is unlikely that the film would maintain such a large mismatch;
pseudomorphic growth is thus ruled out for the face centered phase of Co.
Therefore, the data point (3.1 ML) was set to the natural lattice
parameter of Co (3.544~\AA).  The lattice mismatch of fcc-Co on Cu(001) is
approximately 2\%, and the critical thickness was found to be 3.4 ML.
Thus, both data points are in the permissible region between the two
extreme cases of the model.
Chen et. al. \cite{Chen} observe in-plane anisotropy for Co/Pt(111) films larger than 3.5 ML,
and perpendicular anisotropy for thinner Co films.
Capping these Co/Pt(111) films with thick enough Ag layers however increased the reorientation
thickness to as far as 7 ML, depending on the thickness of the Ag capping layer.
Lee et al. \cite{Lee} found for the systems Co/Pt(111) and Co/Pd(111)
reorientation thickness from perpendicular to in-plane of 5 ML and 12 ML, respectively.
Sandwiches of Pd(111)/Co/Pd were shown to have perpendicular magnetization up to 13 - 15 ML  \cite{Purcell},
which translates to approximately 5~ML.
For Co/Cu(111), a critical reorientation thickness of 5.5 ML is
reported \cite{Lee}. 
Kohlhepp et. al. explain the strong perpendicular anisotropy of Co films on Pd(111) substrates,
capped with either Fe or Cu,
with the alloying of Co and Pd at the Co/Pd(111) interface \cite{Kohlhepp2}.

However, in this thickness range, a gradual phase transformation from the fcc phase
to a hcp phase takes place \cite{Heinz,Lee}.
It is also possible that a bcc cobalt phase is synthesized \cite{Prinz2} in a particular thickness range,
which is not necessarily stable, however \cite{Liu}.
Cobalt was also one of the first systems where magnetic domain formation was found and related with
magnetic anisotropy \cite{Allenspach,Speckmann,Oepen,Kiesielewski}.

The critical thickness of Ni is shown in the bottom part of Fig. 2.
Ultrathin Ni films are in-plane magnetized, and thick films are
out-of-plane magnetized.
This situation is represented in Fig. 1 by plot b).
So far we have not yet addressed the differences in the interface
anisotropy energy, which arises when different chemical elements are
used as substrates, or even when the surface of a film is capped with a different element,

Table III lists a number of surface and interface anisotropy constants
for various systems and chemical species.
Since these anisotropy constants are specifically given for the coordination numbers and packing densities
of the constituents of the interface, including their chemical species, they cannot simply be adopted
for systems with different coordination numbers etc.

The dashed and solid lines in the bottom plot in Fig. 2 represent actually four lines for the critical
thickness, each with surface anisotropy constants for either Ni/Cu$_3$Au(001) and Ni/Cu(001).
The differences of these curves are so minute, however, that the curves cannot be distinguished anymore.
This holds even for the curves which take the magnetocrystalline
anisotropy into account.
The reported critical thickness for Ni/Cu(001) is 12 ML \cite{Tonner}, in good agreement with
our model.

There are no anisotropy constants data available for the system Ni/Cu$_3$Au(001), but we
can try to find a representative average for this system, built from the linear combination
of data for Ni(111)/UHV, Ni/Au(111) and Ni/Cu(001), which are readily
available (Table III).
We assume that the Cu$_3$Au substrate surface is occupied to 3/4 of Cu
atoms and 1/4 of Au atoms.
For simplicity, we ignore here that the Cu$_3$Au substrate may have a Au-rich segregation profile on
its surface \cite{Mecke}, and that alloying may occur between Ni and Cu or Au.

The Cu(001) surface has two atoms per unit mesh and a lattice parameter
of 3.61~\AA. The number density on the surface is thus 1.535 10$^{19}$
atoms/cm$^{2}$.
The interface anisotropy energy constant for Ni/Cu(001) is thus -0.935
10$^{-4}$~eV/atom.
Taking the 3/4 occupation probability of Cu on Cu$_3$Au(001) into account,
we obtain -0.701 10$^{-4}$~eV/atom.

Nickel does not grow pseudomorphic on Au(111) due to the large lattice
mismatch of 15\%.
Instead, we assume that Ni grows on Au(111) in (001) orientation with
its natural lattice parameter of 3.5238~\AA.
The same procedure yields -0.581 10$^{-4}$~eV/atom for the interface anisotropy energy of
Ni/Au(001), or -0.145 10$^{-4}$~eV/atom with 1/4 occupation of Au.

For Ni/Cu$_{3}$Au(001) we thus obtain an occupation weighted value of -0.846 10$^{-4}$ eV/atom
for the interface anisotropy energy constant.

Based on the same considerations, we yield -1.6107 10$^{-4}$eV/atom for the surface
anisotropy of the Ni film vs. the UHV.

Thus, the surface anisotropy and interface anisotropy of Ni/Cu$_{3}$Au(001) amount together to -2.4572
10$^{-4}$eV/atom.

Eq. (16) neglects the magnetocrystalline anisotropy energy, which has a dependence of the
azimuthal angle $\phi$ and thus makes mathematical expressions less transparent then intended here.

The model does not include contributions to the magnetic anisotropy energy that arise from inhomogeneities
such as magnetic domain walls or steps and kinks on surfaces, or surface roughness.
For thicker films, Eq. (16) therefore becomes less accurate.
However, Eqs. (13)-(16) are quite simple and yet elegant expressions which make our anisotropy model transparent enough for
theoretical and also practical considerations.
It is anticipated that these expressions can be implemented with reasonable effort in other formulae,
or that other formulae can be implemented in our anisotropy model.
For instance, the magnetoelastic anisotropy contribution is based on the assumption, that the lattice
parameters of the film material remain unchanged during film growth. This assumption does not really hold,
because of dislocation formation in the film. Implementation of the force balance between film stress and
the tension of dislocations \cite{Matthews2,Lambert} should improve the present model.
Crystal parameters and anisotropy constants are parameters that depend on the temperature.
Crystallographic phase transitions may take place during temperature changes, in particular when
ultrathin films are concerned. Diffusion and interdiffusion processes may be activated at slightly 
elevated temperatures and thus facilitate alloying of film atoms with substrate atoms, which has influence
on the bandstructure of the film.
Finally, the magnetic anisotropy constants are throughout
temperature dependent and may even undergo changes of their sign, which then manifests in the occurrence of so called Hopkinson maxima in
magnetic susceptibility measurements \cite{Wulfhekel}.
While it is difficult to account for all these instances and circumstances, it should be appreciated
that the Bain path at least partially admits to quantitatively model the magnetic anisotropy behaviour
of ultrathin or very thin films on single crystalline substrates.

\section{Conclusions}
A quantitative model for the dependence of the magnetic anisotropy of thin films was established.
It is based on anisotropy constants, lattice strain, and film thickness and takes volume and surface contributions
into account.
Information about the lattice strain is directly implemented in the model as a function of the lattice mismatch via the
epitaxial Bain path.
At the present stage, the model does not take into account contributions that arise from magnetic domain walls, steps,
or surface and interface roughness.
Also, effects of temperature variation (phase transformations, temperature dependent change of sign of anisotropy constants,
etc.) are not accounted for yet.
But the open design of the present model generally allows implementation of these contributions and further refinement,
on the cost of conceptual transparency, however.
Minimization of the total magnetic anisotropy energy of the film yields a closed and simple expression for the critical
reorientation thickness of the magnetic moment of the film.
Experimental data of Fe, Co and Ni films are in fair agreement with our model. 
The significance of magnetoelastic effects on the magnetic anisotropy of thin films is quantitatively verified.
To account for the influence of magnetoelasticity, assessment of elastic strain in the films is indispensable.
A complete structural characterization of the film is thus desirable, which permits to assign the correct crystallographic
phase as well as information about elastic strain.

\begin{acknowledgments}
This work was in part carried out in 1995/96 at Institut
f\"ur Grenz\-fl\"achenforschung and Vakuumphysik (IGV), Forschungszentrum
J\"ulich GmbH, Germany, in the research group of Prof. Matthias Wuttig.
I am grateful to Dr. Benedikt Feldmann (IGV) who encouraged me to develop
the model.
\end{acknowledgments}


\begin{thebibliography}{100}

\bibitem{Bruno}
P. Bruno, 24. IFF Ferienkurs 1993, Forschungszentrum J\"ulich GmbH, J\"ulich, Germany, ISBN 3-89336-110-3.   

\bibitem{Braun}
A. Braun, B. Feldmann, and M. Wuttig, J. Magn. Magn. Mater. {\bf 171} 16 (1997).

\bibitem{4}
G. Bochi, C. A. Ballentine, H. E. Inglefield, S. S. Bogomolov, C. V. Thompson, and
R. C. O'Handley, Mat. Res. Soc. Symp. Proc.
Vol. {\bf 313} 309 (1993).

\bibitem{48}
C. Chappert and P. Bruno, J. Appl. Phys. {\bf 64} 5736 (1988).

\bibitem{1}
P. Bruno and J.-P. Renard, Appl. Phys. A {\bf 49} 499 (1989).

\bibitem{Jezek}
J. Jezek and F. Hrouda, Physics and Chemistry of the Earth {\bf 27} 1247-1252 (2002).

\bibitem{Sander}
D. Sander, Rep.Prog.Phys.  {\bf 62} 809-858 (1999).

\bibitem{Borradaile}
G.J. Borradaile and B. Henry, Earth Sci. Rev. {\bf 42} 49 (1997).

\bibitem{6}
P.M. Marcus and F. Jona, Surface Review and Letters, {\bf 1} (1) 15 (1994).

\bibitem{Alippi}
P. Alippi, P.M. Marcus, and M. Scheffer, Phys. Rev. Lett. {\bf 78} 3892 (1997).

\bibitem{Marcus2}
P.M. Marcus, Surface Review and Letters {\bf 5} 983 (1998).

\bibitem{Feldmann}
B. Feldmann,  Ph.D. Thesis, RWTH Aachen, (1996).

\bibitem{BraunBain}
A. Braun, Surface Review and Letters {\bf 10}(6) 889-895 (2003). 

\bibitem{72}
G.A. Prinz, J. Magn. Magn. Mater. {\bf 100}, 469-480 (1991).

\bibitem{73}
Landolt-B\"ornstein, Bd. III/6, Kap. 1.1.2 1-40 (1989).

\bibitem{Gump}
J. Gump, Hua xia, M. Chirita, R. Sooryakumar, M.A. Tomaz, and G.R. Harp, J. Appl. Phys. {\bf 86}
(11) 6005-6009 (1999).

\bibitem{Heinz}
K. Heinz, S. M\"uller, and L. Hammer, J. Phys.: Condens.  Matter {\bf 11} 9437-9454 (1999).

\bibitem{Escudier}
P. Escudier, Ann. Phys. (Paris) {\bf 9} 125 (1975).

\bibitem{Rebouillat}
J.P. Rebouillat, IEEE Trans. Magn. {\bf 8} 630 (1972).

\bibitem{Paige}
D.M. Paige, B. Szpunar, and B.K. Tanner, J. Magn. Magn. Mat. {\bf 44} 239 (1984).

\bibitem{Gersdorf}
R. Gersdorf, Ph.D. Thesis, Amsterdam (1961), unpublished.

\bibitem{Hubert}
A. Hubert, W. Unger, and J. Kranz, Z. Phys. {\bf 224}, 148 (1969).

\bibitem{Bower}
D.I. Bower, Proc. Roy. Soc. {\bf A326}, 87 (1971).

\bibitem{Kimura}
R. Kimura and K. Ohno, Sci. Rep. Tohoku Univ, {\bf 23}, 359 (1934);
H.J. McSkimin, J. Appl. Phys. {\bf 26}, 406 (1955);
R.M. Bozorth, W.P. Mason, J.H. McSkimin, and J.G. Walker, Phys. Rev. {\bf 75}, 1965 (1949).

\bibitem{51}
B. Heinrich, Z. Celinski, J.F. Cochran, A.S. Arrott, and K. Myrtle, J. Appl. Phys. {\bf 70} 5769 (1991).

\bibitem{52}
T. Takahata, S. Araki, and T. Shinjo, J. Magn. Magn. Mater. {\bf 82}, 287 (1989).
\bibitem{53}
F.J. Lamelas, C.H. Lee, Hui Le, W. Vavra, and R. Clarke,  Mater. Res. Soc. Symp. Proc. {\bf 151}, 283 (1989).
\bibitem{54}
G.H.O. Daalderop, P.J. Kelly, and F.J.A. den Broeder, Phys. Rev. Lett. {\bf 68} 682 (1992).
\bibitem{55}
F.J.A. den Broeder, E. Janssen, W. Hoving and W.B. Zeper, IEEE Trans. Magn. {\bf 28} 2760 (1992).
\bibitem{56}
P.J.H. Bloemen, W.J.M. de Jonge, and F.J.A. den Broeder, J. Appl. Phys. {\bf 72} 4840 (1992).
\bibitem{57}
U. Gradmann, J. Magn. Magn. Mater. {\bf 54-57}, 733 (1986); H.J. Elmers and U. Gradmann,
J. Appl. Phys. {\bf 63}, 3664 (1988).

\bibitem{58}
J.R. Childress, C.L. Chien, and A.F. Jankowski, Phys. Rev. B {\bf 45} 2855 (1992).
\bibitem{59}
E.M. Gregory, J.F. Dillon, Jr., D.B. McWhan, L.W. Rupp, Jr., and L.R. Testardi, Phys. Rev. Lett. {\bf 45} 57 (1980).

\bibitem{60}
G. Xiao and C.L. Chien, J. Appl. Phys. {\bf 61} 4061 (1987).
\bibitem{61}
S. Onishi, M. Weinert, and A.J. Freeman, Phys. Rev. B {\bf 30} 36 (1984).
\bibitem{62}
J. Tersoff and L.M. Falicov, Phys. Rev. B {\bf 26} 6186 (1982).

\bibitem{Matthews2}
J.E. Matthews and A.E. Blakeslee, J. Cryst. Growth {\bf 27}, 118 (1974).

\bibitem{Lambert}
A. Braun, K.M. Briggs, and P. B\"oni,
J. Cryst. Growth  {\bf 241} (1/2) 231 (2002).

\bibitem{Mueller}
S. M\"uller, P. Bayer, C. Reischl, K. Heinz, B. Feldmann, H. Zillgen, and M. Wuttig, Phys. Rev. B {\bf 74} (5) 765 (1995).

\bibitem{Feldmann3}
B. Feldmann, B.Schirmer, A. Sokoll, and M. Wuttig, Phys.  Rev.  B {\bf 57 } (2) 1014-1023 (1998) .

\bibitem{Qiu}
Z.Q. Qiu, J. Pearson, and S.D. Bader, Phys. Rev. B {\bf 48} (13) 8797 (1994).

\bibitem{Thomas}
O. Thomas, Q. Shen, P. Schieffer, N. Tournerie, and B. L\'epine. Phys. Rev. Lett. {\bf 90} (1) 017205 (2003).

\bibitem{Matthews}
J. Tersoff and L.M. Falicov, Phys. Rev. B {\bf 26} 6186 (1982).
J.W. Matthews and J.L. Crawford, Thin Solid Films {\bf 5} 187 (1970).

\bibitem{Chen}
F.C. Chen, Y.E. Wu, C.W. Su, and C.S. Shern, Phys. Rev. B {\bf 66} 184417 (2002).

\bibitem{Lee}
J.-W. Lee, J.-R. Jeong, S.-C. Shin, J. Kim, and S.-K. Kim, Phys. Rev. B {\bf 66}, 172409 (2002).

\bibitem{Purcell}
S.T. Purcell, M.T. Johnson, N.W.E. McGee, W.B. Zeper, and W. Hoving, J. Magn. Magn. Mater. {\bf 113}, 257 (1992).

\bibitem{Kohlhepp2}
J.T. Kohlhepp, G.J. Strijkers, H. Wieldraaijer, and W.J.M. de Jonge,
phys. stat. sol. (a) {\bf 189} (3) 701 (2002).

\bibitem{Prinz2}
G.A. Prinz, Phys. Rev. Lett. {\bf 54} 1051 (1985).

\bibitem{Liu}
A.Y. Liu and D.J. Singh, Phys. Rev. B {\bf 47} (14) 8515 (1993).

\bibitem{Allenspach}
R. Allenspach, M. Stampanoni, and A. Bischof, Phys. Rev. Lett. {\bf 65} 3344 (1990).

\bibitem{Speckmann}
M. Speckmann, H.P. Oepen, and H. Ibach, Phys. Rev. Lett. {\bf 75} 2035 (1995).

\bibitem{Oepen}
H.P. Oepen. M. Speckmann, Y. Millev, and J. Kirschner, Phys. Rev. B {\bf 55} 2751 (1997).

\bibitem{Kiesielewski}
M. Kiesielewski, Z. Kurant, A. Maziewski, M. Tekielak, N. Spiridis, 
and J. Korecki, phys. stat. sol. (a) {\bf 189} (3) 929 (2002).

\bibitem{Tonner}
W.L. O'Brien and B.P. Tonner, Phys. Rev. B {\bf 49} (21) 15370 (1994).

\bibitem{Mecke}
K.R. Mecke and S. Friedrich, Phys. Rev. B {\bf 52} 2107 (1995).

\bibitem{Kohlhepp}
J.T. Kohlhepp and U. Gradmann, J. Magn. Magn. Mater. {\bf 139}, 347 (1995).

\bibitem{Wulfhekel}
W. Wulfhekel, S. Knappmann, B. Gehring, and H.P. Oepen, Phys. Rev. B {50} 16074 (1994).


\end{thebibliography}
\end{document}